\documentclass{agujournal2019}
\usepackage{url} 
\usepackage{lineno}
\usepackage{gensymb}
\usepackage{overpic}
\usepackage{transparent}
\journalname{JGR: Earth Surface}
\nolinenumbers

\begin{document}

\title{Impacts of the cryosphere and atmosphere on observed microseisms generated in the Southern Ocean}



\authors{Ross J. Turner\affil{1,2}, Martin Gal\affil{1,3}, Mark A. Hemer\affil{4}, and Anya M. Reading\affil{1,2}}


\affiliation{1}{School of Natural Sciences, University of Tasmania, Private Bag 37, Hobart, TAS 7001, Australia}
\affiliation{2}{Institute for Marine and Antarctic Studies, University of Tasmania, Private Bag 129, Hobart, TAS 7004, Australia}
\affiliation{3}{Institute of Mine Seismology, 50 Huntingfield Avenue, Huntingfield, TAS 7055, Australia}
\affiliation{4}{CSIRO Oceans and Atmosphere Flagship, Hobart, TAS 7004, Australia}


\correspondingauthor{R. J. Turner}{turner.rj@icloud.com}



\begin{keypoints}
\item Seasonal sea ice coverage is the dominant control on the microseism intensity in East Antarctica
\item Microseism events with extremal intensity are most prevalent in East Antarctica during March-April
\item East Antarctic extremal microseism events increase during the austral autumn for a negative SAM climate index
\end{keypoints}

%
%


\begin{abstract}

The Southern Ocean (in the region 60-180\degree E) south of the Indian Ocean, Australia and the West Pacific is noted for the frequent occurrence and severity of its storms. These storms give rise to high-amplitude secondary microseisms from sources including the deep ocean regions, and primary microseisms where the swells impinge on submarine topographic features. A better understanding of the varying microseism wavefield enables improvements to seismic imaging, and development of proxy observables to complement sparse in-situ wave observations and hindcast models of the global ocean wave climate. We analyse 12-26 years of seismic data from 11 seismic stations either on the East Antarctic coast or sited in the Indian Ocean, Australia and New Zealand. The power spectral density of the seismic wavefield is calculated to explore how the time-changing microseism intensity varies with: i) sea ice coverage surrounding Antarctica; and ii) the Southern Annular Mode (SAM) climate index. Variations in sea ice extent are found to be the dominant control on the microseism intensity at Antarctic stations, which exhibit a seasonal pattern phase-shifted by 4-5 months compared to stations in other continents. Peaks in extremal intensity at East Antarctic stations occur in March-April, with the highest peaks for secondary microseisms occurring during negative SAM events. This relationship between microseism intensity and the SAM index is opposite to that observed on the Antarctic Peninsula. This work informs the complexity of microseism amplitudes in the Southern Hemisphere and assists ongoing interdisciplinary investigations of interannual variability and long-term trends.

\end{abstract}

%
%

%


%
%
%
%

\section{Introduction}
Microseisms are ubiquitously observed on seismometers around the world and arise due to ocean swell and wave activity coupling with the solid earth. Previously considered as `noise' in the study of earthquakes and other seismic events, the usefulness of seismic ambient wavefield \cite{Nakata+2019-CUP} is now recognised since it provides a relatively high frequency energy source for Earth imaging \cite{Shapiro+2005-Science, Pilia+2015-GondRes, Shen+2018-JGR}. Seismic noise caused by Earth surface or oceanic phenomena also have the potential to provide an independent data stream to progress the understanding of the given system and its changes. Examples of the use of ambient seismic noise include volcano monitoring \cite{Sens+2011-ComptesRendusGeo}, the study of stream flow  \cite{Tsai+2012-GRL} and glacier activity \cite{Winberry+2013-GRL}, and very significant progress in understanding and using the relationship between microseisms and ocean waves \cite{Ardhuin+2012-JGR, Obrebski+2012-GRL}.

Microseisms generated by ocean sources are described in terms of the observed frequency \cite{Ardhuin+2019-CUP}.  The longest period signals ($\sim$50-500 s), known as `hum', are generated through the interaction of atmosphere, ocean and sea floor \cite{Rhie+2004-Nature,Traer+2012-JGR}. Primary microseisms ($\sim$10-30 s) are generated through interaction with the sea bed, notably at a continental slope \cite{Hasselmann+1963-RevGeophys}. Secondary microseisms ($\sim$1-10 s) are generated through the interaction of two different wave groups in the deep ocean \cite{Longuet+1950-PhilTransA}.  The alternate name `double frequency' may be understood through the interaction of opposing wave groups of similar frequency \cite{Ardhuin+2015-GRL}.  Recent work has further developed the understanding of microseism generation mechanisms and also the way in which energy from the water column is coupled to the solid-Earth \cite{Gualtieri+2015-JGR, Ardhuin+2018-GRL}.  Microseism studies have been carried out using observatory records from single three-component stations \cite{Aster+2010-GRL} and also using array recordings which afford the ability to locate multiple, simultaneous sources of secondary microseisms \cite{Gal+2016-GJI, Gal+2017-JGR}.  Extending the array analysis to three components where suitable recording exists has helped to reveal the complexity of the ambient wavefield; e.g. the previously undetected \emph{Sn} phase \cite{Gal+2016-GJI}.

\begin{figure}
\centering
\centerline{\includegraphics[width=0.75\columnwidth]{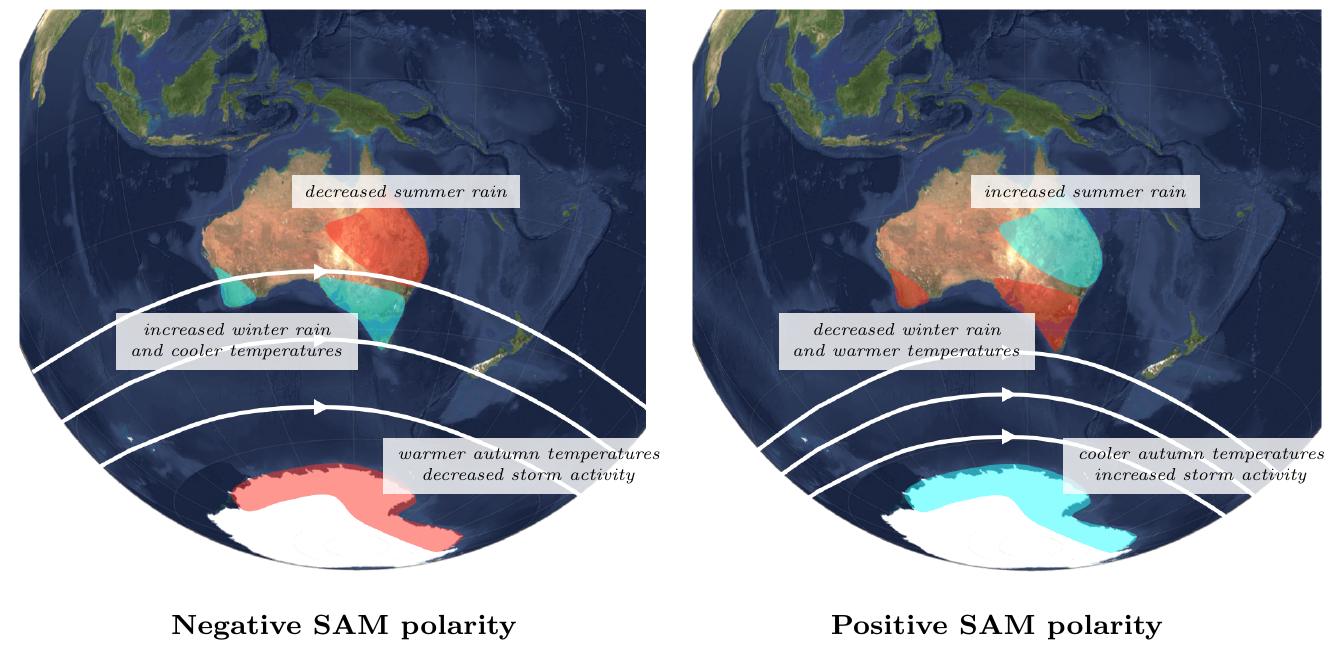}}
\caption{Seasonal impacts of the Southern Annular Mode (SAM) on East Antarctica and Australia. The solid white lines are an approximate location of the circumpolar westerly winds. Regions shaded in red have warmer temperatures, lower precipitation and reduced storm activity; regions in blue have cooler temperatures, increased precipitation and heightened storm activity. The peak season for these anomalies due to variations in the SAM is stated in the overlaid text.}
\label{fig:SAM}
\end{figure}

Where seismic ambient signals are recorded over longer, preferably multidecadal time frames, the ambient wavefield also provides a means of investigating Earth system components that impact the generation of microseisms \cite{Stutzmann+2009-G3, Aster+2010-GRL, Gal+2015-JGR}. Seismic records are now being utilised to improve the homogeneity of ocean wave model hindcasts \cite{Stopa+2019-JGR}, particularly for the Southern Hemisphere where other observations which might be used for hindcast calibration are sparse \cite{Babanin+2019-FMS}.  Cyclical features of global climate are of particular interest, for example the Southern Annular Mode (SAM) climate index, or Antarctic Oscillation (AO), which is associated with variability in wave power \cite{Bromirski+2013-JGROceans, Bromirski+2015-JGROceans}. The SAM concerns the north-south movement of westerly winds that influence the strength and position of cold fronts and mid-latitude systems. SAM indices are defined using either the first empirical orthogonal function (EOF) of a Southern Hemisphere extratropical climate variable (e.g. geopotential height and mean sea level pressure), or the difference in zonal mean pressure between 40 and 65\degree S \cite{Ho+2012-HESS}. The positive mode is associated with the intensification and contraction of the westerly wind belt towards Antarctica resulting in increased polar storm activity \cite{Hurrell+1994-Tellus, Meehl+1998-Tellus, Thompson+2000-JClim, Marshall+2017-GRL}.
Meanwhile, the wind belt expands towards the equator in the negative polarity \cite{Marshall+2007-IJClim, Marshall+2017-GRL} leading to increased storm activity over the Australian continent
\cite{Hemer+2010-IJClim, Marshall+2016-JGR, Wandres+2018-ClimDyn}. The seasonal variations in the SAM are summarised in Figure \ref{fig:SAM}. The SAM has a relatively short characteristic timescale of around 10 days, but the frequency of the positive polarity also has increased slowly on decadal timescales due to stratospheric ozone depletion \cite{Thompson+2011-NatureGeo}. The SAM is recognised to influence global ocean surface waves \cite{Hemer+2010-IJClim, Marshall+2018-OceanMod} and surface temperatures \cite{Marshall+2007-IJClim} which in turn affect sea ice and ice shelves \cite{Greene+2017-SciAdv, Massom+2018-Nature}.

Although noted in passing by \citeA{Aster+2008}, the seasonal variability in microseisms was first investigated at a single station (Dumont d'Urville) as part of a global study undertaken by \citeA{Stutzmann+2009-G3}. Subsequently, the microseism wavefield has been investigated in more detail as part of continent-wide appraisals by both \citeA{Grob+2011} and \citeA{Anthony+2014-SRL}. They reported systematic noise level variations for stations with different installation types and also noted the impact of sea ice in reducing winter noise levels and hence phase-shifting the usual seasonal pattern seen at seismic observatories globally. \citeA{Gal+2015-JGR} investigated patterns of secondary microseism generation including the Southern Ocean using seismic array techniques and found a strong frequency dependency, likely related to water column depth, and preferred locations as observed by the Australian seismic arrays.  In a study focussed on the microseism variations observed from the Antarctic Peninsula, \citeA{Anthony+2017-JGR} looked at the correlation of microseism intensity with sea ice and atmospheric oscillations for the Antarctic Peninsula and Weddell Sea areas of West Antarctica. 

In this contribution, we investigate the influence of seasonal sea ice and sub-seasonal atmosphere/ocean system oscillations on the amplitude of microseisms recorded at permanent observatory stations in the Indian Ocean, Australian and West Pacific sectors of the Southern Ocean.  We aim to add to the work of Anthony et al. \citeyear{Anthony+2017-JGR} noted above, extending the knowledge of typical Southern Ocean microseism variations to eastern latitudes, and hence inform the use of microseisms for Earth imaging and oceanographic applications in the Southern Hemisphere.        

\section{Data}

\subsection{Seismic data}
\begin{figure}
\centerline{\includegraphics[width=0.75\columnwidth]{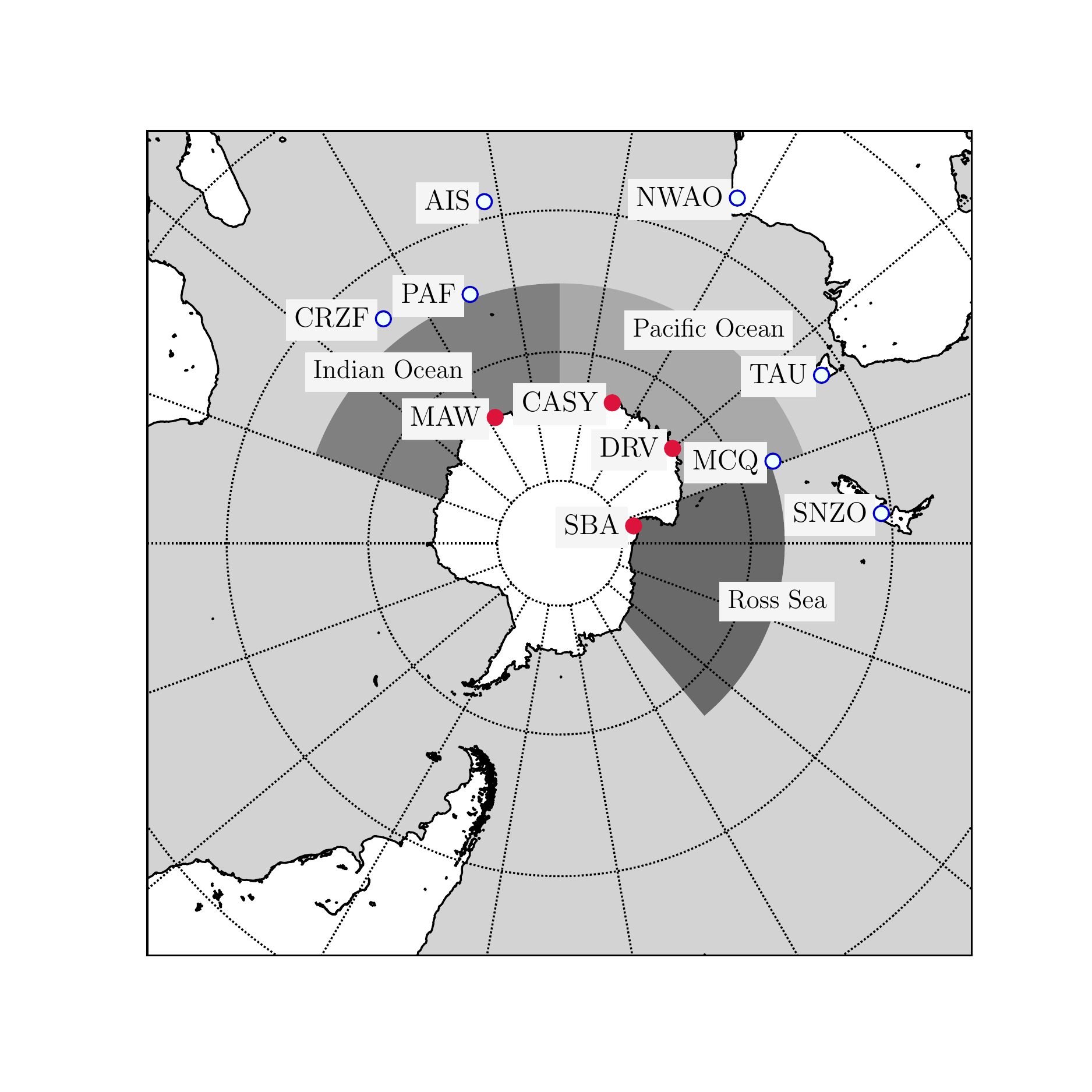}}
\caption{Locations of stations used in this study: East Antarctica (red filled points) and Southern Ocean (blue unfilled points). The geographic definitions of segments within the Southern Ocean (Indian Ocean, Pacific Ocean and Ross Sea) are those used by the National Snow and Ice Data Centre and are shown in shades of grey.}
\label{fig:stations}
\end{figure}

We consider seismic data recorded over a 12-26 year period through to 1 January 2018 at East Antarctic and Southern Ocean stations in the Australian National Seismograph Network (AU), GEOSCOPE (G), and the IRIS/IDA (II) and IRIS/USGS (IU) components of the Global Seismographic Network. Stations were selected based on their locations in the Eastern Hemisphere at latitudes below 30\degree S, and greater than 10 years of data being available for analysis, resulting in a total of 4 Antarctic stations and 7 Southern Ocean stations (Figure \ref{fig:stations} and Table \ref{tab:stations}). We use the vertical component, long-period $1\rm\, Hz$ sampling rate record (LHZ) at each station except for the two seismographs in the Geoscience Australia network which use the broad-band $20\rm\, Hz$ vertical record (BHZ).

\begin{table*}
\caption{Station information, channels and start time of seismic records used in this analysis.}
\small
\centering
\begin{tabular}{l c c c c c c c}
\hline
 Location  & Network & Code & Channel & Latitude & Longitude & Elevation & Start Date \\
 & & & & (degrees) & (degrees) & (m) & \\
\hline
  Casey Station  & IU & CASY & LHZ & -66.279 & 110.535 & 10 & 1996-02-19\\
  Dumont d'Urville  & G & DRV & LHZ & -66.665 & 140.002 & 40 & 1986-02-01\\
  Mawson Station  & AU & MAW& BHZ & -67.604 & 62.871 & 12 & 2003-01-31\\
  Scott Base  & IU & SBA & LHZ & -77.849 & 166.757 & 50 & 1998-10-28\\
\hline
  Crozet Islands  & G & CRZF & LHZ & -46.431 & 51.855 & 140 & 1986-02-01\\
  Hobart  & II & TAU & LHZ & -42.910 & 147.320 & 132 & 1994-01-17\\
  Macquarie Island  & AU & MCQ & BHZ & -54.499 & 158.956 & 14 & 2004-06-28\\
  Narrogin  & IU & NWAO & LHZ & -32.928 & 117.239 & 380 & 1991-11-25\\
  Nouvelle-Amsterdam  & G & AIS & LHZ & -37.796 & 77.569 & 35.9 & 1993-12-25\\
  Port aux Fran\c{c}ais  & G & PAF & LHZ & -49.351 & 70.211 & 17 & 1983-01-01\\
  Wellington  & IU & SNZO & LHZ & -41.309 & 174.704 & 120 & 1992-04-07\\
\hline
\end{tabular}
\label{tab:stations}
\end{table*}

\subsection{Microseism intensity}
\label{sec:Microseism intensity}

\begin{figure}
\centerline{\includegraphics[width=0.8\columnwidth,trim={0 10 0 30},clip]{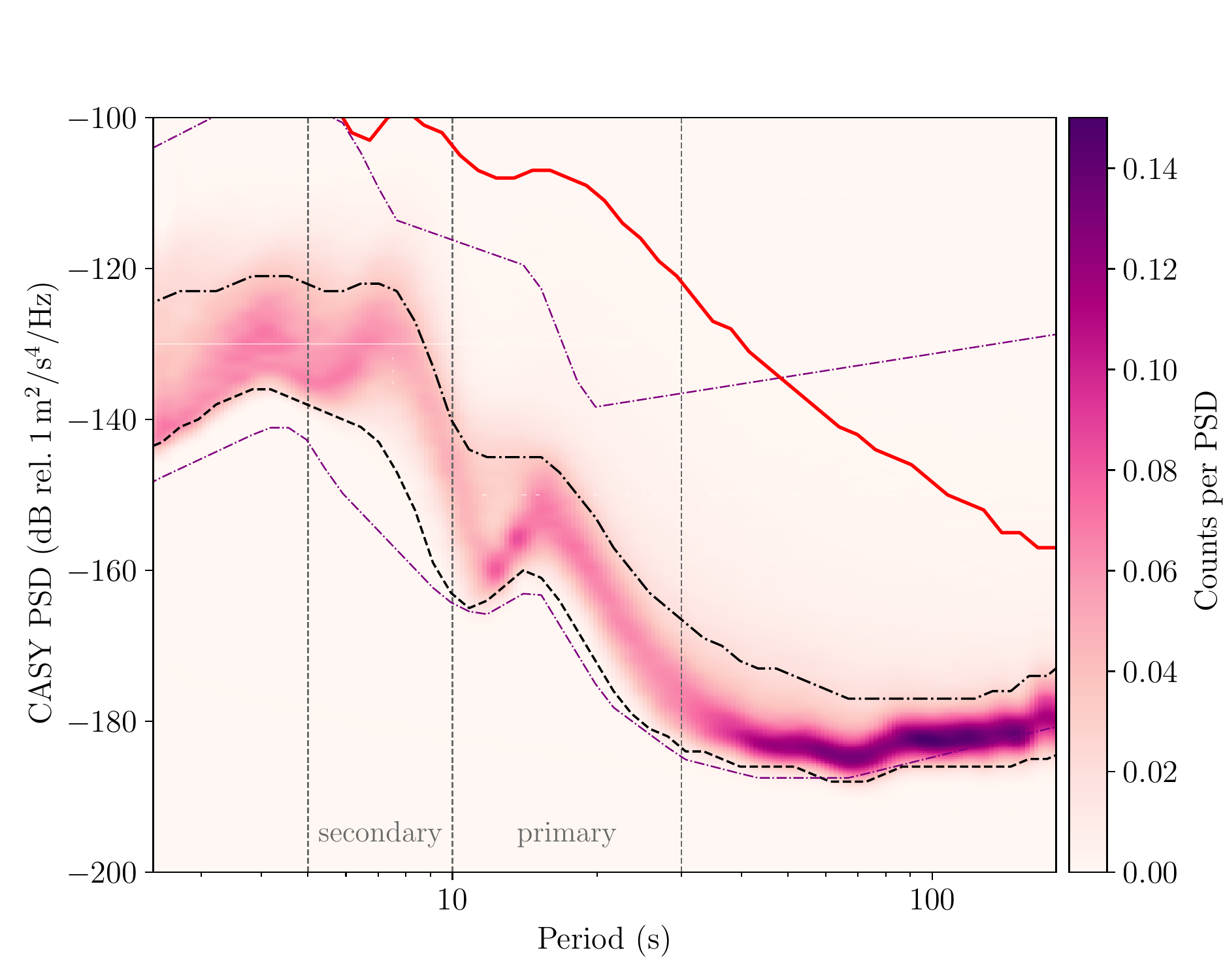}}
\caption{PSD probability density function for Casey Station (CASY), East Antarctica, shown as a time-averaged plot from 1 January 2006 to 1 January 2018. The colour bar shows the average number of counts per PSD in each period bin, gridded in steps of $1\,\rm dB$ along the vertical axis. The 1st and 80th percentiles of the probability density function are shown by the dashed and dot-dashed black lines respectively. The grey dashed vertical lines show the limits of the primary and secondary microseism bands. For comparision, we also plot the \citeA{Peterson+1993} New Low Noise Model and New High Noise Model curves in purple (narrow dot-dashed lines). The solid red line shows the PSD of a magnitude 5.8 earthquake that occurred approximately 100 kilometres south of Casey Station at 20:35 UTC on 4 November 2007.}
\label{fig:spectrum}
\end{figure}

Time series and response files for each of the 11 stations were downloaded from the online archives of the Incorporated Research Institutions for Seismology (IRIS) data services. The continuous time series data were split into three hour windows with 50\% time overlap with each window then being converted to power spectral density (PSD) using the freely available IRIS Noise ToolKit, NTK \cite{IRIS+2014, Hutko+2017-SRL}. The NTK PSD returns power estimates in 1/8-octave period bins and $1\rm\, dB$ power intervals similar to the commonly used PQLX algorithm \cite{McNamara+2011-USGSrep}, and gives the user freedom to choose the smoothing window. Following the method of \citeA{Anthony+2017-JGR}, we specify a 1/4-octave smoothing window to minimise the smearing of power across the period bins. This procedure yields 16 daily measurements of the power spectrum at each station for a minimum of twelve years up to 1 January 2018 (i.e. $>$75000 PSDs per station). The PSD probability density function \cite{McNamara+2004} is shown for CASY station over the time period from 1 January 2006 to 1 January 2018 (Figure \ref{fig:spectrum}).


The broad power spectrum output by the NTK is subdivided for this analysis into two bands containing the energy from the primary (11 - 30 s) and secondary (5 - 10 s) microseisms. The intensity of the primary and secondary microseisms at each time step is calculated by integrating the power spectral density across each of these bands; here, we make use of the IRIS NTK microseism energy bundle \cite{IRIS+2015}.  Earthquakes are flagged and removed following the method of \citeA{Aster+2010-GRL}; the process used to flag earthquakes is detailed in Section \ref{sec:Extremal microseism events}. The microseism intensity then has a median filter applied over a 7 day window with 50\% overlap to ensure the effect of one-off seismic events not directly removed are excluded from the analysis. We flag and exclude data which lies greater than 11 dB outside the mean of the series; this number was chosen by visual inspection to remove only clearly erroneous points. The full time series of the primary and secondary microseism intensity at each station is available in the supplement to the main text.

\begin{figure}
\centerline{\includegraphics[width=0.75\columnwidth,trim={20 45 40 80},clip]{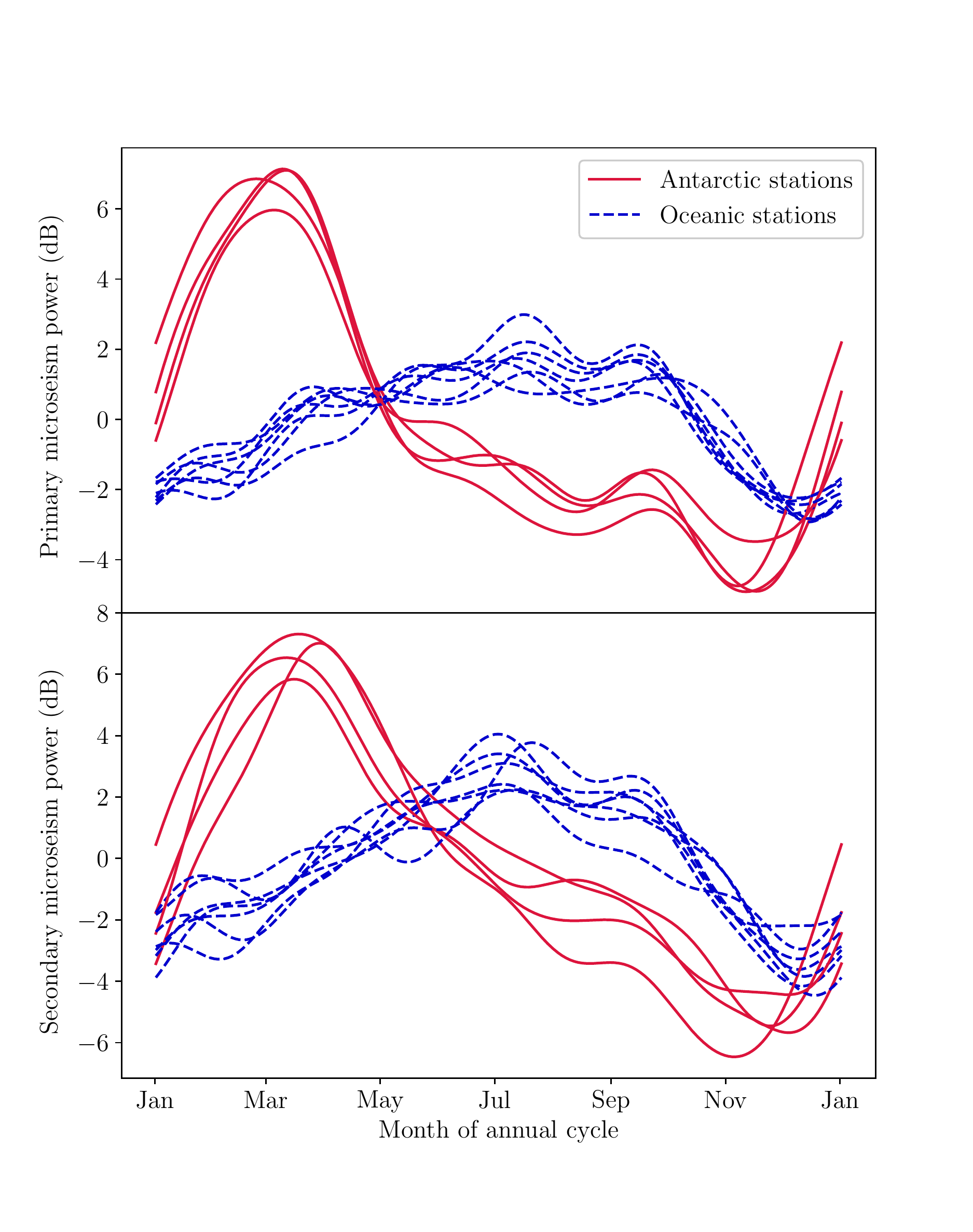}}
\caption{Fourier series fits to the annual cycle in the primary (top panel) and secondary (bottom panel) microseism measured at the four East Antarctic stations (solid red) and seven Southern Ocean stations (dashed blue) between 1 January 2006 and 1 January 2018. The intensity of the primary (and secondary) microseism differs between stations; to enable easy comparison the annual cycles from all stations are shifted to have a mean of zero decibels. The unscaled annual cycles are shown in the supplementary material.}
\label{fig:annualSM}
\end{figure}

The annual periodicity and seasonality of the microseism intensity at each station (Figure \ref{fig:annualSM}) is assessed by fitting a Fourier series to the most recent twelve years of data (due to reliable measurements available at Macquarie Island only after late-2005), following \citeA{Aster+2008}. The Fourier series is fitted to the log-scale intensity assuming a fundamental mode of period one year, a temporal shift in the origin of up to one period, and including both odd and even terms up to sixth-order. The fraction of the variation in the microseism data explained by a periodic annual cycle is measured as the statistic $r^2 = 1 - SSR/SST$, where $SSR$ is the sum of squared differences between the measured microseism intensities and the Fourier series fit, and $SST$ is the sum of squared differences between the measured intensities and the mean intensity. The East Antarctic stations typically have $r^2 \sim 73\%$ of their signal explained by the best-fit periodic annual cycle (median for both primary and secondary bands; see Table \ref{tab:fourier_fits}) when applying a longer 14 day median filter, compared to around 50\% for the Southern Ocean seismographs. Diurnal cycles and short timescale variability that is not cyclical in nature are largely excluded by our choice of the 14 day median filter; the $r^2$ statistic reduces appreciably if this source of variability is included though remains unchanged for longer median filters. Departures from the best-fit curves likely result from moderate to long timescale climate cycles leading to variability in the signal between years (in either amplitude or time of year). 

\subsection{East Antarctic sea ice concentrations}
\label{sec:East Antarctic sea ice measurements}

Sea ice concentrations estimated from satellite microwave backscatter observations were obtained for the Southern Hemisphere from the National Snow and Ice Data Center, NSIDC \cite{Cavalieri+1996-NSIDC}. Daily concentration measurements were retrieved for $25\times25\rm\, km$ grid cells at all southern latitudes below $39$\degree S for the twelve years prior to 1 January 2018 (or between 1 January 1992 and 1 January 2018 for DRV, see Section \ref{sec:Spatial correlation with microseism intensity}).  The polar stereographic projection used by the NSIDC has a nominal grid resolution set at 70 degrees latitude to minimise distortion in the marginal ice zone \cite{Snyder+1987-USGPOrep}; we apply a correction when determining the area of the cells. We assume that a cell covered by at least 15\% sea ice has a damping effect on microseism activity from storms. However, some care must be taken when interpreting these measurements as satellite sensors are known to misreport summertime surface melt as open water leading to underestimates of the concentration. Masks are applied to divide the sea ice concentration observations into geographic regions (see Figure \ref{fig:stations}) and to remove ice-shelves and continental Antarctica from our analysis \cite{Parkinson+2012-Cryo}. In addition, the area of the Southern Pacific Ocean, Southern Indian Ocean and the Ross Sea covered with at least a 15\% sea ice concentration is calculated for each day between 2006 and 2017. We apply a median filter with a 7 day window and 50\% overlap to the sea ice concentration values. This removes any spurious values and provides consistency with the method used for the microseism measurements.

Microseism intensity is expected to have an inverse relationship to sea ice coverage due to the sea ice shielding the Antarctic continent from the influence of ocean storms. The sea ice coverage in each region of the Southern Ocean is normalised so that at the (average) summer minimum the sea ice coverage is 0 and at the (average) winter maximum the sea ice coverage is 1; this scaling is only applied to enable easy comparison between different oceanic regions (raw measurements are shown in the supplementary material). The average annual cycle of sea ice coverage in each region is defined for the purpose of the current study by fitting a Fourier series to the data as for the microseism intensity.

\subsection{SAM index data}
Daily values of the SAM index were retrieved from the National Weather Service Climate Prediction Center \cite{Mo+2000-JClim}. Their SAM index is defined as the leading EOF of the mean monthly 700 hPa height anomalies in the Southern Hemisphere poleward of 20 degrees; daily values are found by projecting short timescale anomalies onto the fitted EOFs. When comparing to the microseism intensities, we apply a median filter with a 7 day window and 50\% overlap to the SAM index to characterise the average behaviour of the SAM over the timescale of the microseism measurements.

\section{Methods}
\subsection{Correlation of sea ice concentration with microseism intensity and SAM Index}
\label{sec:Spatial correlation with microseism intensity}

\begin{figure}
\centerline{\includegraphics[width=1\columnwidth,trim={0 140 0 120},clip]{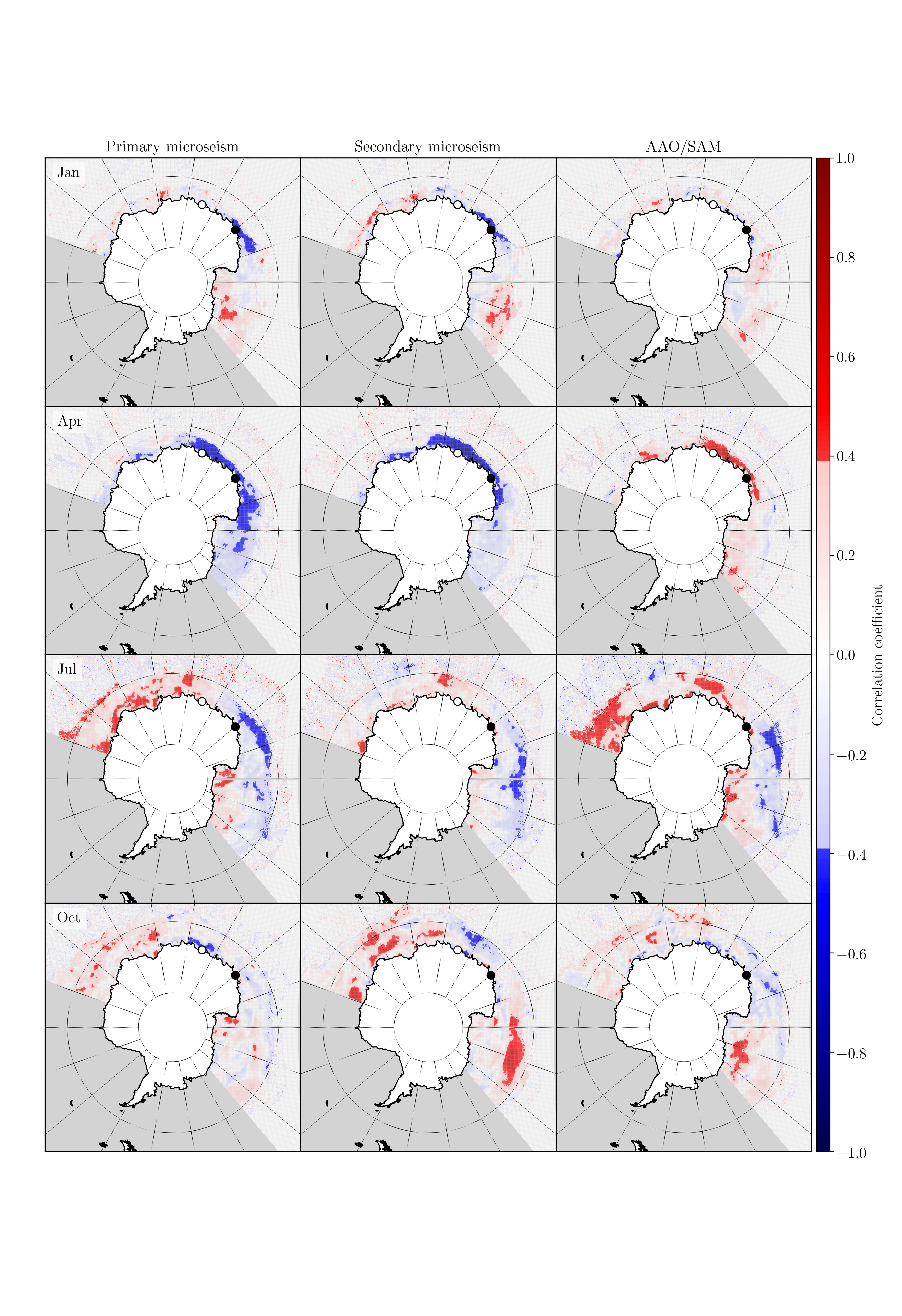}}
\caption{Correlation of monthly average sea ice concentration around East Antarctica; with monthly primary microseism intensity for station DRV (left), monthly secondary microseism intensity for DRV (centre), and SAM index (right). Microseism values are calculated for 26 years of seismic records between 1992 and 2018. Spearman rank correlation coefficients $|r| > 0.39$ are statistically significant at the $2\sigma$ level for our analysis over this time frame.  Correlations are shown as monthly averages for January, April, July and October. The location of DRV is shown using a black filled circle; the location of CASY is shown with an unfilled circle for comparison.}
\label{fig:correlations}
\end{figure}

The relationship between sea ice concentration in a given cell and the microseism intensity as observed at the longest operating East Antarctic seismograph, DRV, is assessed by examining their correlation over a 26 year period. Individual years with extreme values in sea ice concentration are expected to have extreme values in the microseism intensity. This analysis is performed using the sea ice concentrations in each $25\times25\rm\, km$ grid cell in the Southern Pacific Ocean, Southern Indian Ocean and the Ross Sea regions. Following the method of \citeA{Anthony+2017-JGR}, the median sea ice concentration is calculated for each of the twelve 30.44 day months in the calendar years from 1 January 1992 to 1 January 2018. Vectors of length $n = 26$ years are created for each month for each sea ice grid cell. Similarly, the median primary and secondary microseism intensity at station DRV over this 26 year period is grouped into vectors for the twelve 30.44 day long months. The correlation between the sea ice concentration in each cell and the microseism intensity at Dumont d'Urville is calculated by applying the Spearman rank correlation coefficient to the vectors for each month (this statistic considers only the relative order of extreme months). This correlation analysis is repeated for each sea ice grid cell to produce a map of the correlation for each month, as shown in Figure \ref{fig:correlations} for the primary and second microseism intensity. Spearman rank coefficients with an absolute value greater than 0.39 are statistically significant at the $2\sigma$ level for the 26 independent years of observations used in the analysis. 

The correlation of sea ice concentration with the SAM index, associated with the concentration of mid-latitude wind belts towards Antarctica in its positive polarity, is also shown for the same range of years as the microseism intensity given above. The median SAM index is calculated for each 30.44 day long month and the correlation with each of the sea ice grid cells measured (right-hand column of Figure \ref{fig:correlations}). However, the characteristic timescale of the SAM index is less than one month, and so many months may have periods of both positive and negative SAM phase. The 26 years from 1992 to 2018 are ranked for each calendar month based on the ratio of time spent in the positive polarity to that in the negative SAM phase; i.e. the highest/lowest ranked years for a given month will be those exclusively in the positive/negative phase, whilst months with short timescale variations will have 50th percentile rankings. In this manner, we use the Spearman rank coefficient to assess the relationship between the average monthly sea ice concentration and variations in the SAM index on both short and long-term timescales.

\subsection{Correlation of extremal microseism events with SAM index}
\label{sec:Extremal microseism events}

The index hour technique employed by \citeA{Aster+2010-GRL} is applied to the eleven East Antarctic and Southern Ocean stations to identify extremal ocean wave or microseism events, as described below. Firstly, earthquakes must be flagged as these would otherwise comprise the majority of extremal events in our time series. The microseism spectral amplitudes generated by ocean waves decline rapidly at periods longer than 25 seconds \cite{Bromirski+1999-JGR} whereas the signature from earthquakes can be seen to longer periods (see Figure \ref{fig:spectrum}). Earthquake dominated spectra are therefore removed from the time series at each station based upon the spectrum at periods $>$30 seconds. PSDs with 95\% of period bins above 30 seconds exceeding the 80th time-averaged PSD percentile are flagged and removed to produce an earthquake-culled data set \cite{McNamara+2009-SRL}. These selection parameters are tailored to a region with relatively low local seismicity. 

Extremal ocean wave or microseism events are found in the earthquake-culled time series where the 95th percentile in the integrated microseism power is exceeded for three or more contiguous 3 hour PSD estimates (i.e. continuous 6 hours). The time and duration of these events is catalogued for each station (i.e. number of index hours). The earthquake-culled time series is similarly searched for availability of measurements comprising three or more contiguous 3 hour PSD estimates to provide a control level. The extremal microseism index in a given interval (e.g. a month) is then defined as 
the fraction of 6 hour windows that have extremal activity; i.e. the ratio of the duration of times identified as extremal events to the duration of all times forming at least a contiguous 6 hour block. The expected value of the extremal microseism index is thus $< 0.05$ (assuming the 95th percentile cut), though the mean annual value of the index is normalised to unity in this work for clarity. 

\begin{figure}
\centerline{\includegraphics[width=0.9\columnwidth,trim={40 25 50 45},clip]{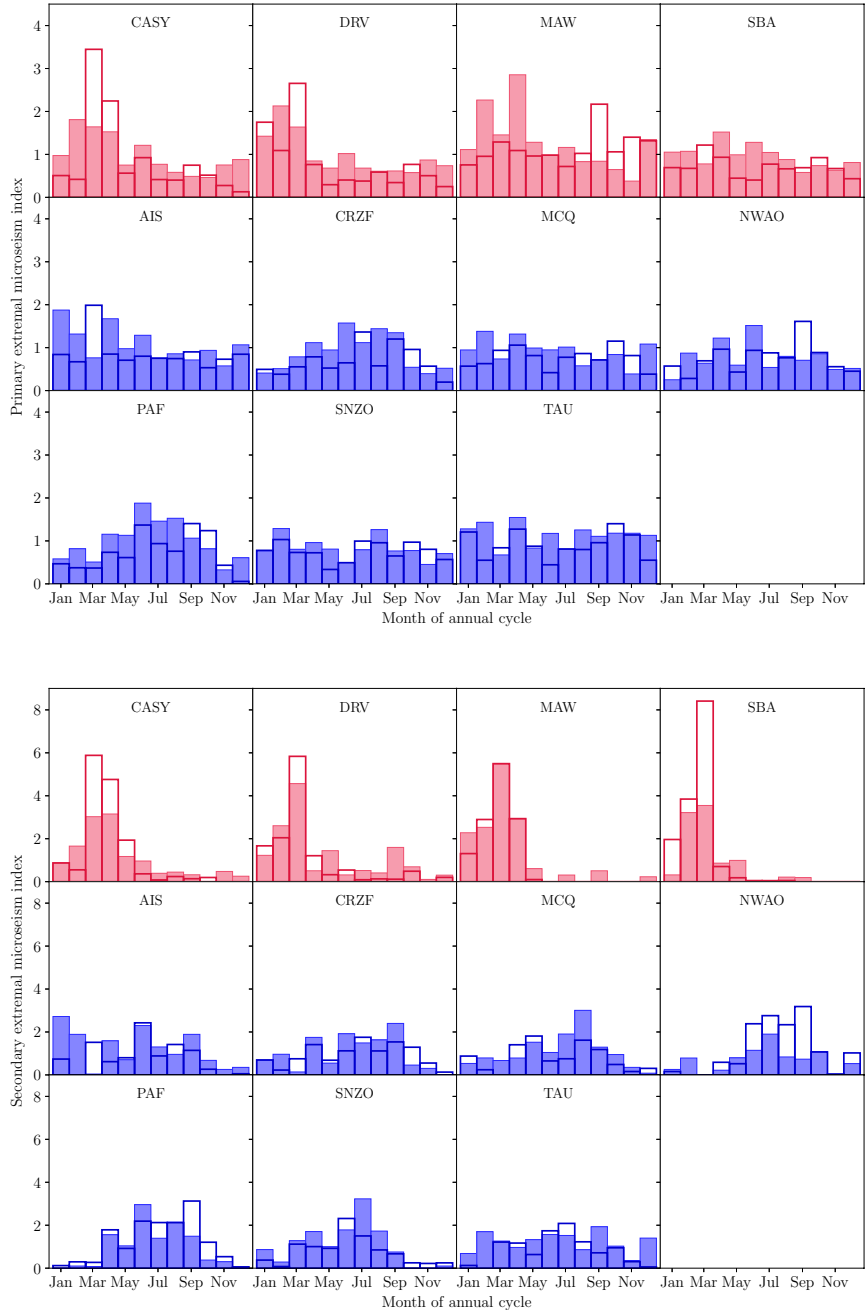}}
\caption{Extremal microseism indices (see text for explanation), for 12 years of observations, binned into 30.44 day months and grouped based upon the value of the SAM in the corresponding 7 day window. Extremal microseism indices for negative SAM values (below the 30th percentile) are shown as unfilled bars. Microseism indices for positive SAM values (above the 70th percentile) are shown as shaded bars. The extremal microseism indices in the top panel are calculated using the primary microseism and normalised so that duration of all storms in an average month is unity. The indices in the bottom panel are calculated for the secondary microseism.}
\label{fig:storm_indices}
\end{figure}

The extremal primary and secondary microseism indices are further grouped based upon the value of the Southern Annular Mode (SAM). The twelve year period from 1 January 2006 through 1 January 2018 is subdivided into 30.44 day long months and the median SAM index for each month calculated. Months with an index below the 30th percentile are assumed to be predominantly in the negative phase and those with an index above the 70th percentile are in the positive phase. The mean extremal primary and secondary microseism indices are thus calculated for each of the twelve 30.44 day long months comprising a calendar year, but including only the months classified as in either the positive or negative SAM phase (Figure \ref{fig:storm_indices}).

\section{Results}
\label{sec:res}

\subsection{Microseism intensity}

The component of the microseism intensity explained by a periodic annual cycle is fitted by a sixth-order Fourier series (as explained previously) for each of the eleven East Antarctic and Southern Ocean stations. The Fourier fits to the secondary microseism intensity are shown in Figure \ref{fig:annualSM} and are grouped based upon their geographic location (i.e. East Antarctica or Southern Ocean). The secondary microseism intensity at all East Antarctic stations increases 6-10 decibels above the time-averaged intensity during March then slowly drops off before bottoming at 4-6 decibels below the average during November. In contrast, the seven oceanic stations peak in July with a smaller 2-3 decibels above the time-averaged intensity and bottom out 2-3 decibels below the average in December. The fluctuation from the mean value of the primary microseism intensity follows a similar pattern to that of the secondary microseism across all stations.


The East Antarctic stations also show a greater degree of annual periodicity than those in the Southern Ocean based on the fraction of the microseism intensity explained by the Fourier fits (Table \ref{tab:fourier_fits}). Specifically, a median of 74\% of the variation in the East Antarctic stations is due to an annual cycle compared to 47\% at the oceanic stations, with the remainder resulting from sub-seasonal and interannual variations (see Figures S1 and S2 in the supporting information). The March peak in the microseism intensity at the four Antarctic stations is confidently detected above short timescale signals superimposed on the annual cycle in each of the 12 years examined in this work (Figure S8); the amplitude of this peak shows only minor interannual variations (e.g. no unusually quiet or noisy years). The July peak in oceanic stations is similarly higher than the level in December each year (Figures S9 and S10). The primary microseism intensity shows a slightly greater degree of annual periodicity than the secondary microseism for seven of the eleven stations though the difference is not statistically significant.

\begin{table}
\caption{The fraction of the variation in the microseism intensity explained by a periodic annual cycle, as fitted using a high-order Fourier series. The first column is the name of the station, the second and fourth the day of the fitted annual cycle with maximum microseism intensity in the primary and secondary bands respectively, and the third and fifth columns are the corresponding $r^2$ statistics of the Fourier fits.}
\small
\centering
\begin{tabular}{l c c c c}
\hline
 Seismograph  & \multicolumn{2}{c}{Primary microseism} & \multicolumn{2}{c}{Secondary microseism} \\
   & peak day & $r^2$ & peak day & $r^2$  \\
\hline
  Casey Station  & March 12 & 0.73 & March 29 & 0.70 \\
  Dumont d'Urville  & February 24 & 0.67 & March 16 & 0.54 \\
  Mawson Station  & March 10 & 0.80 & March 18 & 0.76 \\
  Scott Base  & March 5 & 0.74 & March 12 & 0.79 \\
\hline
  Crozet Islands  & July 12 & 0.51 & July 3 & 0.55 \\
  Hobart  & July 20 & 0.43 & July 13 & 0.34 \\
  Macquarie Island  & June 9 & 0.49 & July 6 & 0.46 \\
  Narrogin  & July 18 & 0.43 & July 21 & 0.60 \\
  Nouvelle-Amsterdam  & July 16 & 0.53 & July 3 & 0.49 \\
  Port aux Fran\c{c}ais  & July 18 & 0.59 & July 7 & 0.63 \\
  Wellington  & June 28 & 0.46 & July 5 & 0.36 \\
\hline
\end{tabular}
\label{tab:fourier_fits}
\end{table}

The increase in microseism intensity during the southern winter at the seven oceanic stations is explained by seasonal storm activity. The East Antarctic stations meanwhile are shielded by sea ice throughout the winter which is expected to dampen storm activity and thus the microseism intensity \cite{Grob+2011}. The correlation between sea ice and microseism intensity is explored further in the next section.

\subsection{Sea ice extent and microseism intensity correlation}
\label{sec:Sea ice extent and microseism intensity correlation}

\begin{figure}
\centerline{\includegraphics[width=0.75\columnwidth,trim={0 120 0 140},clip]{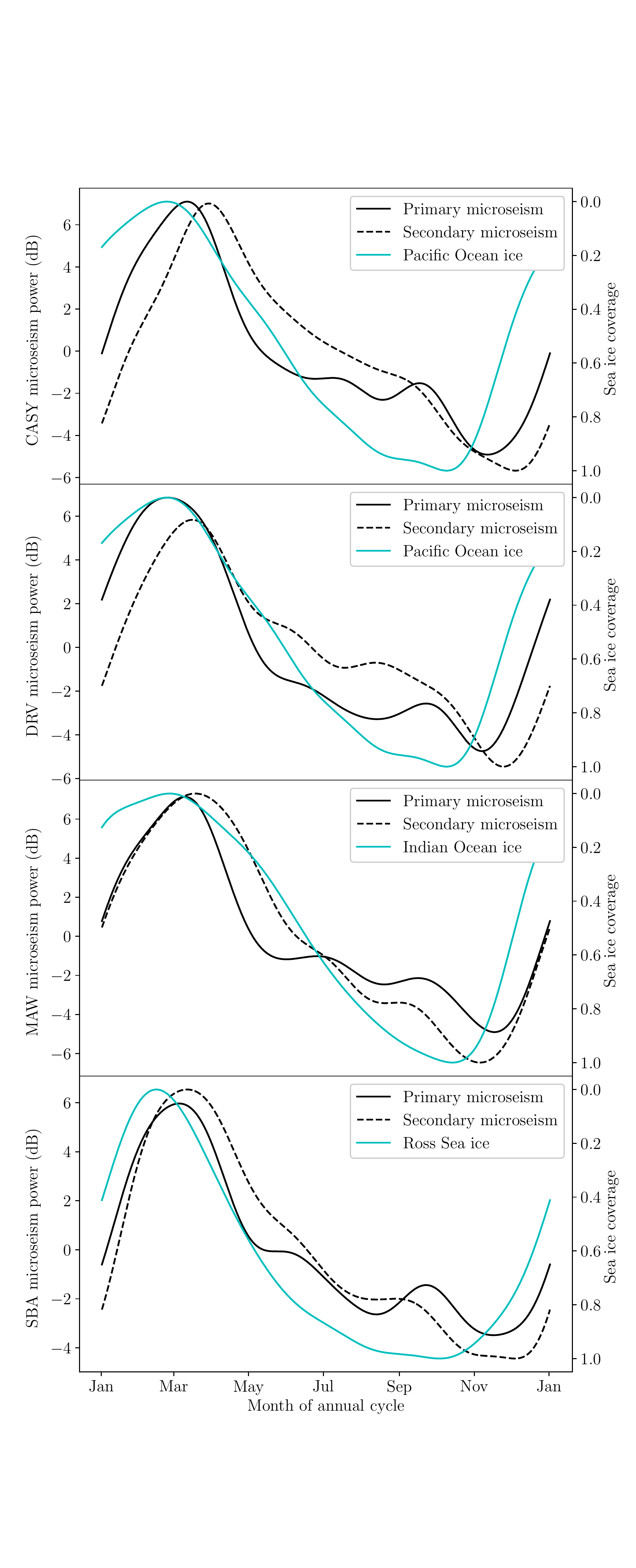}}
\caption{Fourier series fits to the annual cycle in the primary (solid black) and secondary (dashed black) microseism measured at the four Antarctic stations between 1 January 2006 and 1 January 2018. The intensity of the primary and secondary bands differ by a few orders of magnitude; to enable easy comparison these are both shifted to have a mean of zero decibels. The Fourier series fit to the sea ice observations in the region of the Southern Ocean surrounding the relevant station are plotted in cyan for comparison. Note that the sea ice coverage axis is reversed to explore the potential causal relationship with microseism intensity.}
\label{fig:sea ice}
\end{figure}

Sea ice is expected to impede the generation of both the primary and secondary microseisms at Antarctic stations through the damping and continental shelf shielding of ocean waves (preventing the formation of microseisms in proximity to the coast). 
The annual periodicity in the microseism intensities at Casey Station is compared to the annual cycle in the area of the South Pacific Ocean exposed through ice melt in the top panel of Figure \ref{fig:sea ice} (see Figure S3 in the supporting information for fit to the raw data). The area of ocean covered by sea ice is plotted on an inverted axis to more easily examine its correlation with the microseism intensity, noting these variables are expected to be negatively correlated for Antarctic stations. The annual cycle in the inverted sea ice coverage has a comparable shape to the microseism intensity (in log units/decibels), with the minimum ice coverage occurring in February and the maximum coverage in October. The peak in the primary microseism intensity lags the inverted sea ice coverage by approximately a month whilst the peak in the secondary microseism lags the primary by a further month. A similar pattern is observed at Dumont d'Urville, Mawson Station and Scott Base as shown in Figure \ref{fig:sea ice}.

The fraction of the annual cycle in the primary and secondary microseisms at each station explained by seasonal sea ice melt, and the time lag between these processes, is investigated by fitting the inverted sea ice coverage Fourier series to the microseism intensity Fourier series. The amplitude, phase (or time lag) and vertical offset of the sea ice Fourier series are free parameters in these fits. The time lags that best align the annual cycles in the primary and secondary microseism intensities to the cycles in the (inverted) sea ice coverage at the four East Antarctic stations are shown in Table \ref{tab:seaice_corr}; the $r^2$ statistic of the fits is also provided. The primary microseism lags the annual sea ice melt cycle by between 10 and 27 days whilst the secondary microseism lags the ice melt by between 21 and 56 days. Variation in time lags between stations is due to their disparate geographic locations; e.g. the sea ice may impinge on the continental shelf which is responsible for generating the primary microseism at different times near each station. The sea ice melt cycle leads the annual cycle in the primary and secondary microseism intensity at greater than the $15\sigma$ level at each East Antarctic station. The annual cycle in the primary microseism intensity further leads the cycle in the secondary microseism, significant at least at the $5\sigma$ level. These results support the hypothesis that increased sea ice coverage leads to suppressed microseism intensities in coastal East Antarctica; i.e. as expected, the annual cycles in sea ice extent and microseism intensity are inversely related, with a fitted annual phase shift of between 125 and 170 degrees (4-5 months). Moreover, sea ice coverage explains between 78 and 93\% of the variation in the annual cycle in the primary microseism intensity and between 85 and 95\% of the variation for the secondary microseism. 

Maximum microseism intensities lagging the summer sea ice minimum may be explained by strengthening ocean waves increasing storm activity in the austral autumn. Meanwhile, the lag between the winter sea ice maximum coverage and minimum microseism intensity is likely due to a general decline in southern ocean storminess from October to December \cite{Young+1999-IJC}. The relative importance of the sea ice and ocean wave state controls on the microseism intensity in East Antarctica are compared. We assume that the average Fourier series for the microseism intensity (separately for the primary and secondary bands) across the seven oceanic stations is representative of the microseism signal generated by the ocean wave climate. This proxy for the annual cycle in the ocean wave climate is fitted to the microseism intensities at the East Antarctic stations as before, though assuming zero phase shift. The wave climate proxy explains 13\% of the variation in the annual cycle of the secondary microseism intensity at Casey Station, with much smaller fractions for the primary microseism and at the other stations. The seasonal variation in sea ice extent is therefore the dominant control on the microseism intensity in East Antarctica.

\begin{table*}
\caption{Time lag and correlation between measurements of the primary and secondary microseism compared to the fractional sea ice coverage for the four East Antarctic stations. The first column is the name of the seismograph, the second the corresponding region of ocean, the third and fifth the delay between the sea ice observations and the primary and secondary microseisms respectively (the inverted sea ice coverage leads the increase in microseism intensity), and the fourth and sixth are the $r^2$ statistic of the correlation.}
\small
\centering
\begin{tabular}{l c c c c c}
\hline
 Seismograph  & Ocean & \multicolumn{2}{c}{Primary microseism} & \multicolumn{2}{c}{Secondary microseism} \\
 & & delay (days) & $r^2$ & delay (days) & $r^2$ \\
\hline
  Casey Station  & Pacific Ocean & $26.9\pm0.6$ & 0.82 & $56.3\pm0.4$ & 0.91 \\
  Dumont d'Urville  & Pacific Ocean & $9.7\pm0.5$ & 0.87 & $41.2\pm0.5$ & 0.85 \\
  Mawson Station  & Indian Ocean & $16.4\pm0.7$ & 0.78 & $20.9\pm0.5$ & 0.92 \\
  Scott Base  & Ross Sea & $15.8\pm0.3$ & 0.93 & $27.6\pm0.3$ & 0.95 \\
\hline
\end{tabular}
\label{tab:seaice_corr}
\end{table*}

\subsection{Seasonality of extremal microseism events}

The seasonal variation in the extremal microseism index is examined for the four East Antarctic and seven Southern Ocean stations in months where the Southern Annular Mode (SAM) is classified as either the positive or negative polarity. The seasonality of the secondary microseism index for the Antarctic stations is shown in the histograms in the bottom panel of Figure \ref{fig:storm_indices} (red bars). The overall pattern seen in the secondary extremal microseism index for East Antarctic stations is driven by the same mechanisms as the microseism intensity previously examined; here we are interested in the difference between the positive and negative phases. In general, the extremal microseism index is higher from March to April during the negative SAM phase but higher from June through December during the positive phase. 
Weak seasonal variations with the SAM are seen in the extremal primary microseism index at the East Antarctic stations, as shown in the top panel of Figure \ref{fig:storm_indices}, however the patterns are not consistent between stations. The number of extremal events in the primary band at Mawson and Scott Base also shows minimal variation throughout the year in stark contrast to the strong annual cycle seen in their microseism intensity. Meanwhile, the extremal microseism indices at the oceanic stations show only a weak relationship with the SAM in both the primary and secondary bands (see Figure \ref{fig:storm_indices}). The secondary microseism index at NWAO (Narrogin, Western Australia) is however notably higher during the negative polarity from June to September. The significance of these findings and a comparison with previous studies in this field is made in the discussion (Section \ref{sec:Discussion}).

The correlation between the extremal microseism index at the four East Antarctic stations on interannual timescales is examined in Figure \ref{fig:EastAntarctic_storms}. The number of extremal events measured at Casey Station and Dumont d'Urville peaks between 2005 and 2009, and is at a minimum between 2011 and 2013. By contrast, the number of extremal events at Mawson Station peaks between 2012 and 2014, whilst Scott Base has a constant level across all years.

\section{Discussion}
\label{sec:Discussion}

The microseism study in this work examines the correlation between sea ice coverage, and the SAM index, with the microseism intensity across East Antarctica and the surrounding Southern Ocean. Similar research has previously been undertaken by \citeA{Anthony+2017-JGR} focused on the northern tip of the Antarctic Peninsula and the Drake Passage, using measurements at Palmer Station (PMSA) and East Falkland Island (EFI) respectively. This previous work found that the primary and secondary microseism intensity at Palmer Station peaks in May, falls until September, rises to a small peak in November, before reaching a minimum (of similar level to September) around January. By contrast, the intensities in East Antarctic peak two months earlier in March and fall steadily until November before increasing again. The shifted peak in the microseism intensity results from the earlier annual sea ice expansion around East Antarctica (e.g. these stations are between 2 and {15\degree} further south). The increasing storm activity into the austral autumn/winter is thus detected at Palmer Station for longer before being suppressed by the growing sea ice. Similarly, the increase in microseism intensity (after the minimum) occurs two months later in East Antarctica than at Palmer Station as the sea ice retreat occurs earlier along the northern tip of the Antarctic Peninsula. The annual pattern in the microseism intensities measured at East Falkland Island is similar to the oceanic stations of our current investigation. \citeA{Anthony+2017-JGR} suggest that the primary microseism at EFI results from Rayleigh waves generated off the west coast of the Antarctic Peninsula and therefore remain somewhat linked to seasonal variations in sea ice extent (the primary microseism intensity between May and September is suppressed); however our oceanic stations show no clear signature of the annual sea ice cycle due to their greater distance from the Antarctic continent. \citeA{Grob+2011} also conducted a initial analysis of the relationship between sea ice extent and microseism intensity at coastal stations across the Antarctic continent obtaining similar results on both the Antarctic Peninsula and East Antarctica.

The secondary microseism signal at East Antarctic stations is found to lag the primary microseism by between 4.5 and 32 days, with the greatest time delays at Casey Station and Dumont d'Urville. The time lag between the annual cycles in the two microseism bands is significant at the $5\sigma$ level. The large time lag is indicative of different source generation regions for the primary and secondary microseisms. Primary microseisms generated through the interaction of ocean waves on the continental shelf are expected to be quickly impeded as sea ice forms. The concentration of near-coastal sea ice surrounding Dumont d'Urville has a significant negative correlation with the primary microseism intensity well into the austral winter (Figure \ref{fig:correlations}); this is consistent with the findings of \citeA{Anthony+2017-JGR} at Palmer station. By contrast, secondary microseisms generated through the interaction of opposing wave groups in the deep ocean are much less sensitive to near-coastal sea ice; the secondary microseism shows minimal correlation with the sea ice concentration beyond the austral autumn. These findings may aid efforts to reconstruct the spatial and temporal distribution of sea ice around Antarctica based on the microseism amplitudes \cite{Cannata+2019}. 

The duration of extremal secondary microseism events measured at the East Antarctic stations is higher in March and April during a negative SAM phase than for the positive polarity. Specifically, at Casey Stations during these two months, the extremal index is between 45\% and 110\% greater during the negative SAM polarity when measured using either the primary or secondary microseism. At Scott Base, the duration of extremal events measured in March using the secondary microseism are 140\% greater during a negative phase than for the positive SAM polarity. This trend is also seen in Dumont d'Urville, though to a lesser degree, whilst Mawson Station (MAW) does not show a consistent pattern for the two microseism bands. However, the spectral amplitudes at MAW are also approximately an order of magnitude lower than the other stations, which \citeA{Grob+2011} suggest may indicate an incorrect instrument response. By contrast, the comparable study of extreme microseism events recorded on the Antarctic peninsula by \citeA{Anthony+2017-JGR} found increased storm activity in the Drake Passage during the positive SAM polarity leads to greatly increased microseism intensity levels (primarily from October to January). During the austral winter, we similarly find an increased number of extremal microseism events in the positive SAM phase.

These measurements of relatively subdued East Antarctic microseismic activity in the austral autumn during positive SAM events are in contrast to the expected intensification of the polar vortex and westerly circumpolar winds, and the corresponding increase in Antarctic storm activity \cite{Hurrell+1994-Tellus, Meehl+1998-Tellus, Thompson+2000-JClim, Marshall+2017-GRL}. This finding may be explained by the different growth rate of the sea ice during the austral autumn for the two SAM polarities. The sea ice concentration around East Antarctica is strongly correlated with the SAM index during April (Figure \ref{fig:correlations}). The reduced sea ice concentration is likely explained by the warmer East Antarctic temperatures during the negative SAM phase \cite{Marshall+2007-IJClim}. During years with a negative SAM polarity in much of the austral autumn, warmer surface temperatures are expected to lead to slower or delayed growth of the sea ice around East Antarctica. The delayed growth of the East Antarctic sea ice thus exposes the coastal East Antarctic stations directly to any storm activity, even though these storms may be weaker or less frequent than during the positive polarity. By contrast, in the Antarctic Peninsula \cite{Marshall+2007-IJClim} the surface temperature is higher for the positive SAM phase; greater storm activity should be detected during the positive polarity as the sea ice retreat is faster or occurs earlier. Meanwhile, during the austral winter the growth of new sea ice shifts deeper into the Southern Ocean; the concentration of sea ice formed closer to the continent in previous months becomes independent of the SAM index due to its short characteristic timescale. The sea ice extent is therefore expected to become independent of the SAM index on the order of one characteristic timescale after the summer minimum (i.e. approximately one month; on longer timescales sea ice growth will have occurred under both polarities).

The extremal microseism indices at Narrogin in Western Australia are also increased during the negative SAM polarity, though in the austral winter/spring during July, August and September. Specifically, the number of extremal index is 130 to 340\% higher during September for the negative SAM phase, and up to 180\% higher in July and August (Figure \ref{fig:storm_indices}). These measurements are consistent with the accepted interpretation that, during the negative SAM polarity, the belt of strong westerly winds expands towards the equator leading to increased storm activity over the Australian continent \cite{Hemer+2010-IJClim, Marshall+2016-JGR, Wandres+2018-ClimDyn}. Port aux Fran\c{c}ais similarly has an increased secondary microseism extremal index over the austral winter during the negative SAM polarity, whilst the winter extremal index at Macquarie Island is instead higher in the positive phase due to its more southern location (i.e. wind belt moves north of MCQ in the negative phase). The extremal index at the other oceanic stations shows no clear relationship with the SAM, perhaps due to the influence of local climate drivers.

\begin{figure}
\centerline{\includegraphics[width=0.75\columnwidth,trim={0 120 0 140},clip]{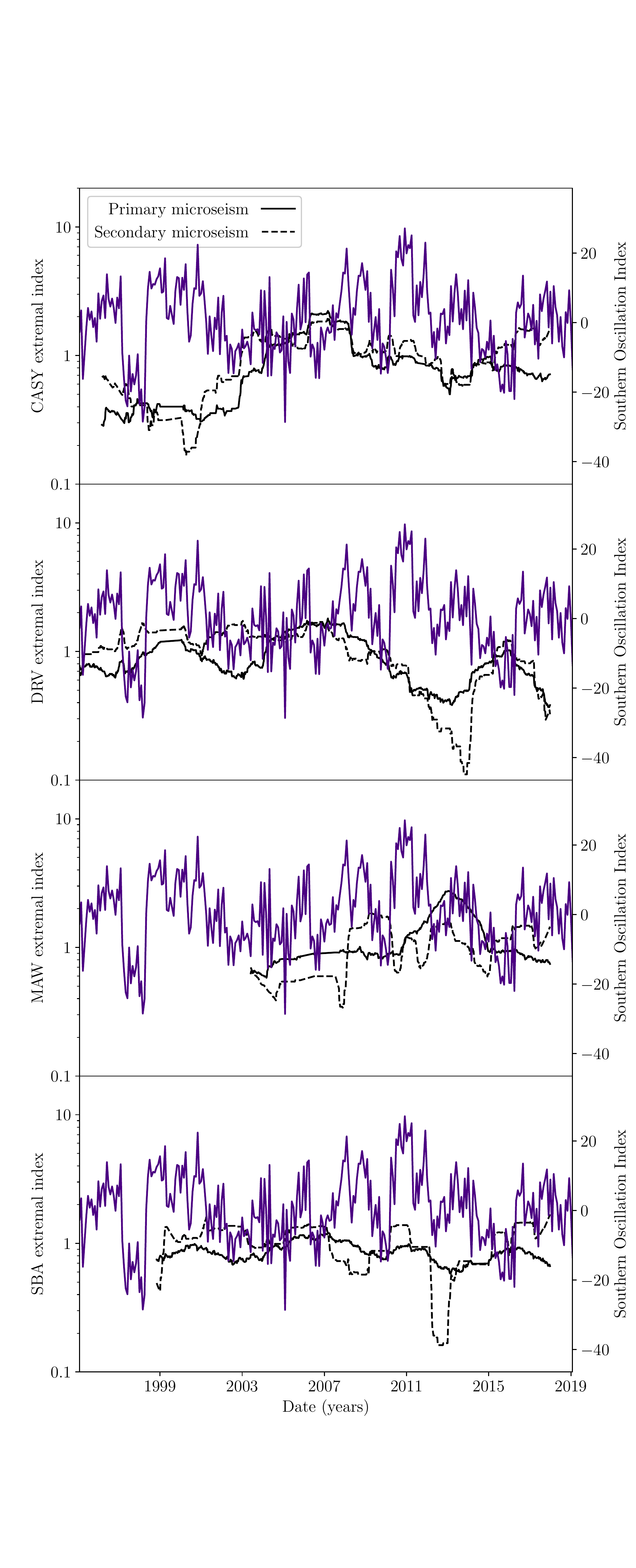}}
\caption{Extremal microseism index measured at the four East Antarctic seismographs for their operational history (or after 1995). The black solid line is the one-year moving average of the primary microseism index and the dashed line is the moving average for the secondary microseism. The El Ni\~{n}o-Southern Oscillation index is plotted on the secondary axis in purple for comparison.}
\label{fig:EastAntarctic_storms}
\end{figure}

We examined the correlation of the extremal activity with other climate indices (that have longer characteristic timescales than SAM) to explain the interannual variability. Specifically, we find no correlation between the microseism index at East Antarctic stations and the El Ni\~{n}o-Southern Oscillation index (shown in Figure \ref{fig:EastAntarctic_storms}) or other widely-used Southern Hemisphere climate indices. Interannual variations in the extremal microseism index (associated with storm activity) may be explained by a combination of several climate indices, including more localised climate variability such as the Pacific South American teleconnection patterns \cite{Marshall+2016-JGR, Marshall+2017-GRL}.

The improved knowledge of the annual cycle of ambient microseism levels will aid the planning of Earth imaging studies in Antarctica. Researchers collecting data for ambient seismic tomography, for example, can ensure that instruments are deployed during months of strong signal levels and understand departures from the omni-directional wavefield assumption. Further, as the impacts of global change appear to be taking place in East Antarctica faster than previously anticipated \cite{Witze+2018-Nature, Fox+2019-Science}, and since both sea ice loss and ocean swell can play a part in ice shelf disintegration \cite{Massom+2018-Nature}, microseism intensity may prove useful in understanding rapid changes in these interacting Earth systems. Recent use of microseism intensity for the detection of long-term climate trends by the oceanography community \cite{Stopa+2019-JGR} may be further developed using the relationships we have identified in this contribution.

\section{Conclusions}

In this contribution, we have analysed 12-26 years of microseism observations from 11 seismographs on the East Antarctic coast, in Australia, New Zealand and the Southern Indian Ocean, to characterise the variability of recorded primary and secondary microseism intensities.  The relationship between satellite observations of sea ice extent and microseism occurrence, and the influence of the Southern Annular Mode (SAM) index were also investigated.

Annual variations in the sea ice extent are the dominant control on the microseism intensity measured at East Antarctic stations, explaining a median of $\sim$73\% of both the primary and secondary band microseism signal. By contrast, the Southern Ocean wave climate explains at most 13\% of the microseism signal in coastal East Antarctica. Consistent with previous studies, the microseism intensity cycle in Antarctica shows a seasonal pattern phase-shifted by 4-5 months compared to seismic stations from other Southern Hemisphere continents, with higher levels observed in the austral summer and autumn months.

We observe a time lag of approximately one month between the minimum sea ice coverage and the maximum primary microseism intensity for East Antarctic stations.  The reforming sea ice quickly curtails the production of primary microseisms with an inferred source region close to the continental shelf.  The lag of a further month between the primary and secondary microseism intensity peaks suggests a deeper ocean source for the secondary microseisms.

Peaks in the number of extremal microseism events for the area of the Southern Ocean investigated in this contribution occur during March and April, with the peaks at Casey Station and Scott Base between 45 and 140\% higher during these months for a negative SAM polarity. Previous work \cite{Anthony+2017-JGR} found the opposite relationship is observed on the Antarctic Peninsula where an increased number of extremal events are detected during the positive SAM phase. This could be explained by the warmer surface temperatures present around coastal East Antarctic during the negative polarity leading to a weakened or delayed growth of sea ice in the early austral autumn; by contrast, the surface temperatures are colder along the Antarctic Peninsula during the negative phase. The storm activity in Western Australia is also strongly anti-correlated with the SAM index; the number of extremal events in September increases by between 130 to 340\% during the negative polarity.

We have found microseism intensity at stations in the Southern Ocean exhibit a region-specific response to variations in the SAM climate index, in contrast to the consistent annual pattern seen across coastal Antarctica due to seasonal sea ice growth. Understanding such controls on microseism intensity will prove valuable in the use of microseisms as a proxy observable for the ocean wave state to study the impacts of global change on East Antarctica. In particular, both sea ice and ocean swell have been observed to foreshadow ice shelf disintegration.
Constraints from seismology, together with in-situ ocean wave observations, can assist in the calibration of satellite data to investigate both interannual variability in the Southern Ocean wave state and any underlying climate trends from ocean wave hindcast reanalyses.

\acknowledgments

This research was supported by the Australian Research Council through the Discovery Program (Project DP150101005) and the Special Research Initiative for Antarctic Gateway Partnership (Project SR140300001). We thank Rob Anthony and an anonymous reviewer for their helpful and constructive feedback that has improved this manuscript.

The facilities of the IRIS Data Management Center were used for access to waveforms and related metadata, used in this study. IRIS Data Services are funded through the Seismological Facilities for the Advancement of Geoscience and EarthScope (SAGE) Proposal of the National Science Foundation under Cooperative Agreement EAR-1261681. This work included data from the AU, G, II, and IU seismic networks obtained from the IRIS data centre; \url{https://doi.org/10.18715/GEOSCOPE.G}, \url{https://doi.org/10.7914/SN/II}, and \url{https://doi.org/10.7914/SN/IU}.

\clearpage

\bibliography{bibliography}

\end{document}